\documentclass[useAMS,usenatbib,aas_macros]{mn2e}
\input{psfig}
\usepackage{color}

\newcommand{\be}{\begin{equation}}
\newcommand{\ee}{\end{equation}}

\def\no{\noindent}

\newcommand{\ifm}[1]{\relax\ifmmode#1\else$\mathsurround=0pt #1$\fi}
\newcommand{\kms}{\ifmmode\,{\rm km}\,{\rm s}^{-1}\else km$\,$s$^{-1}$\fi}

\newcommand{\ltsima}{$\; \buildrel < \over \sim \;$}
\newcommand{\lsim}{\lower.5ex\hbox{\ltsima}}
\newcommand{\gtsima}{$\; \buildrel > \over \sim \;$}
\newcommand{\gsim}{\lower.5ex\hbox{\gtsima}}

\def\M11{M_{11}}
\def\V100{V_{100}}
\def\R1{R_{Mpc}}
\def\T6{T_6}

\title[On the jet contribution to the AGN cosmic energy budget] 
{{On the jet contribution to the AGN cosmic energy budget}} 

\author[A. Cattaneo, P.~N. Best]
{A.~Cattaneo$^{1\star}$, P.~N.~Best$^{2}$\\
\\
$^1$Astrophysikalisches Institut Potsdam, an der Sternwarte 16, 14482 Potsdam, Germany\\
$^2$SUPA, Institute for Astronomy, Royal Observatory, Blackford Hill, Edinburgh EH9 3HJ, UK\\
$^\star$acattaneo@aip.de}
\begin{document}

\pagerange{\pageref{firstpage}--\pageref{lastpage}} \pubyear{2005}

\maketitle

\label{firstpage}


\begin{abstract}
Black holes release energy via the production of photons in their
accretion discs but also via the acceleration of jets. We investigate the
relative importance of these two paths over cosmic time by determining the
mechanical luminosity function (LF) of radio sources and by comparing it
to a previous determination of the bolometric LF of active galactic nuclei
(AGN) from X-ray, optical and infrared observations.  The mechanical LF of
radio sources is computed in two steps: the determination of the
mechanical luminosity as a function of the radio luminosity and its
convolution with the radio LF of radio sources.  Even with the large
uncertainty deriving from the former, we can conclude that the
contribution of jets is unlikely to be much larger than $\sim 10\%$ of the
AGN energy budget at any cosmic epoch.

\end{abstract}



\section{Introduction}
\label{sec:intro}

Matter can accrete onto a black hole (BH) only if it releases a fraction
$\epsilon\sim 0.06-0.4$ of its rest-mass energy, where $\epsilon$ depends
on the BH spin \citep{bardeen70}.  In the standard
\citet{shakura_sunyaev73} model, the energy is dissipated by viscous
torques in the accretion disc and radiated.  Accretion from a luminous
disc provides a physical model to explain the luminosity of quasars
\citep{lynden-bell69}.

Several authors have computed the total energy radiated by BHs over cosmic
time and have compared it to the local BH mass density
\citep{soltan82,chokshi_turner92,yu_tremaine02,barger_etal05,hopkins_etal07}.
The two are in good agreement if matter is turned into light with a
canonical efficiency of $\epsilon\sim 0.1$.  This has been used as an
argument to infer that most of the BH mass in the Universe was accreted
luminously, but it only proves that $>25\%$ of the BH mass in the Universe
was accreted luminously, since $\epsilon$ could be as large as
$\epsilon\sim 0.4$ if most BHs were maximally rotating.

In fact, accreting BHs (`active galactic nuclei', AGN) produce not only
light but also jets of matter, which are radio-luminous because of the
synchrotron radiation from ultrarelativistic electrons accelerated in
shocks.  Excluding objects beamed toward the line of sight, only a small
fraction of the luminosity that is radiated by 
an AGN comes from synchrotron radiation.
However, the synchrotron power represents only a small fraction of the jet
mechanical luminosity, most of which may be used to do work on the
surrounding gas.

Moreover, at low accretion rates ($\dot{M}_\bullet\lsim 0.01\dot{M}_{\rm
Edd}$, where $L_{\rm Edd}=\epsilon\dot{M}_{\rm Edd}{\rm c}^2$ is the
Eddington luminosity), the accretion disc is not dense enough to radiate
efficiently; the disc puffs up, and the energy that needs to be removed to
allow the accretion may be carried out more easily by jets.  Although the
physics of this picture are still speculative, AGN that channel a large fraction of
the accretion power into jets while showing little emission from an accretion disc are observed
(eg \citealp{dimatteo_etal03}; \citealp{allen_etal06}).  These radio
sources are less powerful than quasars but more common due to their longer
duty cycle. For example, around two-thirds of brightest cluster galaxies
are radio galaxies \citep{burns90,best_etal07}.  In contrast, only one
galaxy in $10^4$ contains a quasar ($M_B < -23$) at $z\sim 0$
\citep{wisotzki_etal01}.  Finally, mechanical energy is thermalized in the
intracluster medium more efficiently than luminous energy is.  The
observational evidence that the mechanical heating by AGN is important to
solve the cooling-flow problem in galaxy groups and clusters is getting
strong (\citealp{best_etal05}; \citealp{dunn_fabian06};
\citealp{rafferty_etal06}; \citealp{magliocchetti_brueggen07}; see also
Cattaneo et al. 2009, and references therein).

For these reasons, it is important to compare the mechanical and radiative
output of AGN. This is the goal of the current paper. This issue has also
recently been addressed by
\citet{shankar_etal08a}, \citet{kording_etal08} and \citet{merloni_08}. We
adopt a different approach to these authors and produce results that are
qualitatively similar. 
The layout of our paper is as follows.  In Section~2, we analyse
how we can use radio data to infer the jet mechanical power.  We shall see
that two different approaches give different $P_{\rm jet}(L_{\rm radio})$
relations. We consider both, and use the difference between the results obtained from the two relations
to provide an estimate of the uncertainty.  We convolve these
relations with the radio LF of radio sources, $\phi(L_{\rm radio})$, to
estimate the mechanical LF, $\phi(L_{\rm radio}(P_{\rm jet}))$, first in
the local Universe (Section~3), then at different redshifts (Section~4).
In both cases, we integrate over luminosity to determine the
mechanical power per unit volume, which we compare with the radiative
power per unit volume from luminous AGN.  In Section~5 we summarize the
results of comparing the mechanical LF of AGN to the bolometric luminosity
function of AGN determined by \citet{hopkins_etal07} and we discuss the
implications of our results.

\section{The mechanical luminosity of a radio source}
\label{sec:mechlum}

Obtaining an estimate of the mechanical power of a radio source is an
inherently difficult problem. The observed monochromatic radio luminosity
measures only the fraction of the jet power that is currently being
converted into radiation. That fraction is small (typically between 0.1
and $1\%$; cf. \citealp{bicknell95}) and changes during the lifetime of
the radio source, since the radio luminosity of a growing radio source
first increases and then drops as the source expands into a progressively
lower density environment (eg \citealp{kaiser_etal97}). Nevertheless, it
is reasonable to expect that radio and mechanical luminosities should show
at least a broad degree of correlation on a population basis.

Estimates of the mechanical power of radio sources have followed two
approaches.  The first \citep{willot_etal99} uses the minimum energy
density $u_{\rm min}$ that the plasma in the radio lobes must have in
order to emit the observed synchrotron radiation (eg \citealp{miley80}).
With this approach, the jet mechanical power is $L_{\rm mech}\gsim u_{\rm
min}V/t$, where $V$ is the volume filled by the radio lobes and the
radio-source lifetime $t$ is given by the ratio between the jets' length
and the hotspots' advancement speed.  The largest sources of uncertainties
are: (i) the nature of the jet plasma (electron-positron or
electron-proton?): the value of $u_{\rm min}$ is larger if the lobes
contain a hadronic component in addition to the synchrotron radiating
particles (relativistic electrons and/or positrons); (ii) the lack of
observational constraints on the low-frequency cut-off of the electron
energy distribution: for a synchrotrom spectrum $\propto\nu^\alpha$ with
radio spectral index $\alpha<-0.5$, $u_{\rm min}$ is larger when the lower
cut-off frequency takes a lower value and thus there is more energy in the
synchrotron spectrum.  \citet{willot_etal99} derive the relation

\begin{equation}
L_{\rm mech} = 3 \times 10^{38} f_{\rm W}^{3/2}
    \left(\frac{L_{\rm 151\,MHz}}{10^{28}{\rm W\,Hz}^{-1}{\rm
    sr}^{-1}}\right)^{6/7}{\rm W},
\label{w1}
\end{equation} 

\noindent where $f_W\sim 1-20$ incorporates all the unknown factors.
\citet{blundell_rawlings00} argue for $f_{\rm W} \simeq 10$ for
\citet{fanaroff_riley74} class II sources (FR\,IIs), while
\citet{hardcastle_etal07} suggest $f_{\rm W}= 10-20$ for FR\,Is.  We
convert the luminosity at 151$\,$MHz, $L_{\rm 151\,MHz}$, into a luminosity at 
1.4$\,$GHz, $L_{\rm1.4\,GHz}$, where the local radio LF
is best-determined. This will also allow us later to compare Eq.~(\ref{w1})
with another determination of $L_{\rm mech}$ by Best et al. (2006, 2007).
For the conversion we assume a spectral index $\alpha=-0.8$.  Using
$f_{\rm W}=10$, Eq.~(\ref{w1}) gives:

\begin{equation}
L_{\rm mech} = 1.4 \times 10^{37} \left(\frac{L_{\rm
  1.4\,GHz}}{10^{25}{\rm W\,Hz}^{-1}}\right)^{0.85}{\rm W}.
\label{w2}
\end{equation}

\noindent Notice that $L_{\rm 151\,MHz}$ is given in ${\rm W\,Hz}^{-1}{\rm
sr}^{-1}$ while $L_{\rm 1.4\,GHz}$ is given in ${\rm W\,Hz}^{-1}$ to respect
the different conventions used by \citet{willot_etal99} and
\citet{best_etal06}, so there is a factor of $4\pi$ entering the
conversion.

A second approach is to infer $L_{\rm mech}$ from the mechanical work that
the lobes do on the surrounding hot gas. The expanding lobes of
relativistic synchrotron-emitting plasma open cavities in the ambient
thermal X-ray emitting plasma, which advances in X-ray imaging
capabilities now allow to be imaged in detail. The minimum work in
inflating these cavities is done for reversible (quasi-static) inflation
and equals $pV$, where $p$ is the pressure of the ambient gas.
\citet{best_etal06} derived a relation between radio and mechanical
luminosity based upon this estimate for the energy associated with these
cavities, combined with an estimate of the cavity ages from the buoyancy
timescale (from \citealp{birzan_etal04}).  Comparing the mechanical
luminosities of 19 nearby radio sources that have associated X-ray cavities
with their 1.4\,GHz monochromatic radio luminosities leads to a relation

\begin{equation}
L_{\rm mech} = (3.0 \pm 0.2)\times 10^{36}\, f\, \left(\frac{L_{\rm
1.4\,GHz}}{10^{25}{\rm W\,Hz}^{-1}}\right)^{0.40 \pm 0.13}{\rm W},
\label{b1}
\end{equation} 

\noindent broadly in agreement with that derived by \citet{birzan_etal04}.
In Eq.~(\ref{b1}), the factor $f$, incorporated by \citet{best_etal07},
accounts for any systematic error in estimating the mechanical
luminosities of the cavities.  In particular, $pV$ is likely to be an
underestimate of the energy needed to inflate a cavity: the enthalpy of
the cavity is $\frac{\gamma}{\gamma -1}pV = 4pV$ for the relativistic
plasma in the radio lobes, suggesting that $f\sim 4$ may be appropriate.  Some
authors have even argued for mechanical energies in excess of $10pV$
\citep{nusser_etal06,binney_etal07} due to additional heating directly
from the jets. For $f=4$, Eq.~(\ref{b1}) gives:

\begin{equation}
L_{\rm mech} = 1.2 \times 10^{37} 
\left(\frac{L_{\rm 1.4\,GHz}}{10^{25}{\rm W\,Hz}^{-1}}\right)^{0.40}
{\rm W}.
\label{b2}
\end{equation}
 
Eqs.~\ref{w2} and~\ref{b2} are in excellent agreement at $L_{\rm 1.4\,GHz}
\sim 10^{25}{\rm W\,Hz}^{-1}$, but the relation derived from the
minimum-energy argument has a steeper radio luminosity dependence than the
relation derived from the X-ray cavities. Nevertheless, given the totally
independent approaches used to derive Eqs.~\ref{w2} and~\ref{b2}, and the
uncertainty factors in both equations, the degree of consistency is
encouraging. Indeed, the discrepency between the two relations may
simply reflect the fact that Eq.~\ref{w2} is determined from powerful,
radiatively efficient radio sources (mostly $L_{\rm 1.4\,GHz} \gsim
10^{25}{\rm W\,Hz}^{-1}$) while Eq.~\ref{b2} is derived predominantly from
lower luminosity, radiatively inefficient sources. The mean relation
between $L_{\rm mech}$ and $L_{\rm rad}$ may well be different in these
two different regimes, in which case a combination of Eq.~\ref{w2} at high
luminosities and Eq.~\ref{b2} at low luminosities would be most
appropriate.

A third approach towards estimating mechanical luminosities of radio
sources has been developed by \citet{merloni_08}. These authors use the
de-beamed radio core emission as a measure of the jet kinetic luminosity,
based upon the analogy with X-ray binary sources and the so-called
Fundamental Plane relation for black holes \citep{merloni_etal03}. This
approach uses the radio LF of flat-spectrum (ie core-dominated) radio
sources as a measure of the LF of radio cores. It then requires
assumptions about the statistical de-beaming of radio sources
(ie the distribution of Lorentz factors of jets), and about how to
correct for the radio cores that are missed from the flat-spectrum radio
LF because their radio sources are dominated by extended steep spectrum
emission. These factors can be reasonably estimated for moderate
to high radio luminosity sources in the local Universe, but they are not so well
constrained at low radio luminosities, where the bulk of the jet
mechanical power is produced, nor at higher redshifts. We therefore do not
consider this approach here, but we do compare our results with the results obtained by
\citet{merloni_08} in Section~\ref{sec:concl}.

\begin{figure*}
\noindent
\centerline{\hbox{\psfig{figure=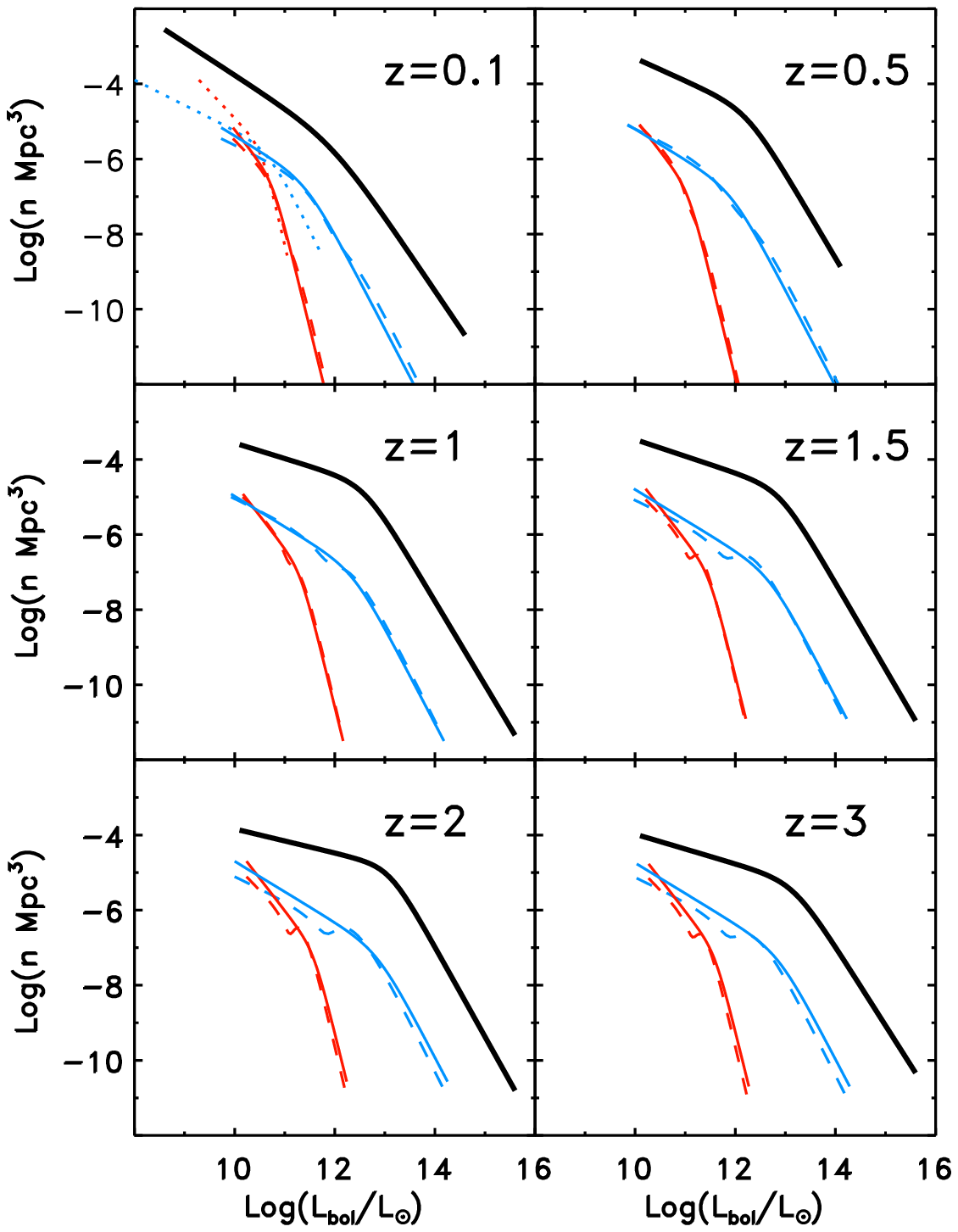,height=20.cm,angle=0}}}
\caption{The thick solid lines show the bolometric LF of AGN inferred by
\citet{hopkins_etal07} from X-ray, optical and infrared data.  The thin
solid and dashed lines show the mechanical LFs inferred from the radio LFs
of \citet{dunlop_peacock90} and \citet{willot_etal01}, respectively.  At
$z\simeq 0$, we have also shown the mechanical LFs inferred from
\citet{best_etal05a}'s local radio LF (dotted lines).  For each radio LF,
we have determined two mechanical LFs, according to the two different
$P_{\rm jet}(L_{\rm radio})$ conversions derived in
Section~\ref{sec:mechlum}: the red lines are for $P_{\rm jet}\propto
L_{\rm radio}^{0.4}$ (Eq.~\ref{b2}) and the blue lines are for $P_{\rm
jet}\propto L_{\rm radio}^{0.85}$ (Eq.~\ref{w2}).}
\label{fig1}
\end{figure*}

\section{The local mechanical LF}

In order to derive a mechanical LF for radio sources in the nearby
Universe, we must convolve Eqs.~\ref{w2} and~\ref{b2} with the local radio
LF. Here we adopt the local 1.4$\,$GHz radio LF of \citet{best_etal05},
which is derived from the Sloan Digital Sky Survey spectroscopic sample
and is fully consistent with other recent determinations of the local
radio LF (eg \citealp{machalski_godlowski00}; \citealp{sadler_etal02}).
Throughout this paper we follow the convention of defining the LF
$\phi(L)$ as the number of objects per unit volume and
logarithmic-luminosity interval (eg \citealp{hopkins_etal07}), so the
number of sources with luminosity between $L$ and $L+{\rm d}L$ is
$(\phi/{\rm ln}10){\rm d}L/L$.  With this definition, the
\citet{best_etal05} LF can be be parameterised using the double power-law
model

\begin{equation}
\phi(L) = \phi_* \left[ \left(\frac{L}{L_*}\right)^\alpha + 
                     \left(\frac{L}{L_*}\right)^\beta \right] ^{-1},
\label{blf}
\end{equation}

\noindent where $L=L_{\rm 1.4\,GHz}$ (we wrote $L$ without any subscripts
because we shall use Eq.~\ref{blf} to model other LFs).  The best-fit
parameters are $\phi_* = 10^{-5.7}$Mpc$^{-3}$, $L_* =
10^{25.16}$W\,Hz$^{-1}$, $\alpha = 0.57$ and $\beta = 2.31$.  Combining
Eq.~(\ref{blf}) with Eqs.~\ref{w2} and~\ref{b2} gives the blue and the red
dotted curves in the $z=0.1$ diagram of Fig.~1, respectively.

The mechanical power that is released per unit volume is

\begin{equation}
\rho_{\rm mech} =\int_0^\infty L_{\rm mech}(L){\phi(L)\over{\rm
    ln}10}{{\rm d}L\over L}
\label{blf2}
\end{equation}
$$=A\left({L\over 10^{25}{\rm\,W}}\right)^\eta\int_0^\infty{{\rm d}x\over
x^{\alpha+1-\eta}+x^{\beta+1-\eta}},$$ 

\noindent where $A=4.8\times 10^{11}L_\odot$ and $\eta=0.85$ for
Eq.~(\ref{w2}), and $A=1.0\times 10^{11}L_\odot$ and $\eta=0.4$ for
Eq.~(\ref{b2}).  If $L_{\rm mech}$ is computed with Eq.~(\ref{w2}), then
$\rho_{\rm mech}^{\rm synchro}=1.6\times 10^5L_\odot{\rm\,Mpc}^{-3}$
($1L_\odot=3.9\times 10^{26}{\rm\,W}$).  If $L_{\rm mech}$ is computed
with Eq.~(\ref{b2}), the problem is more complicated because in that case
the integral in Eq.~(\ref{blf2}) diverges at $x=0$. This is because the
double power-law model in Eq.~(\ref{blf}) cannot be extrapolated down to
$L\rightarrow 0$, as can be easily seen: the faint-end slope of the local
radio LF is steeper than that of both the low luminosity end of the local
galaxy optical LF (eg \citealp{norberg_etal02}) and the low mass end of
the local mass function of supermassive black holes
(eg \citealp{shankar_etal08b}). Therefore, if the radio LF
extrapolated too far, then the calculated space density of radio-loud AGN
would exceed that of galaxies (or supermassive black holes) capable of
hosting them.  For example, if the slope of the radio LF were to remain unaltered
down to $L_{\rm 1.4\,GHz}=10^{17}{\rm W\,Hz}^{-1}$ then the local space
density of radio galaxies would exceed that of galaxies integrated down to
$M_B \sim -15$.

Considering that only massive galaxies with massive black holes have a
significant probability of harbouring a radio source powered by an AGN,
the local space density of radio sources matches that of supermassive
black holes with $M_\bullet \gsim 10^6 M_{\odot}$ 
(ie $\sim 5\times 10^{-3}{\rm\,Mpc}^{-3}$ according to \citealp{shankar_etal08b}) 
if the radio LF is
extrapolated down to $L_{\rm 1.4\,GHz} \sim 10^{19.2}{\rm W\,Hz}^{-1}$. This
sets a strong lower limit on the luminosity to which the radio LF can be
extrapolated, and so for a conservative calculation we evaluate the
integral in Eq.~(\ref{blf2}) by adopting $10^{19.2}{\rm W\,Hz}^{-1}/L_*\simeq
10^{-6}$ as the lower extreme of the integration interval. With this
choice, the mechanical power per unit volume from Eq.~(\ref{b2}),
$\rho_{\rm mech}^{\rm cav}$, is about ten times larger than the
mechanical power per unit volume from Eq.~(\ref{w2}), $\rho_{\rm
mech}^{\rm synchro}$. The discrepency would reduce to a factor of three if
a limit of $10^{22}{\rm W\,Hz}^{-1}$ were adopted instead.

We want to compare these values to the radiative power per unit volume
from luminous AGN.  \citet{hopkins_etal07} made the first attempt at
determining the bolometric LF of AGN by combining hard X-ray, soft X-ray,
optical and infrared data.  They found that the double power-law model in
Eq.~(\ref{blf}) fits their results with $\phi_*$, $L_*$, $\alpha$ and
$\beta$ dependent on redshift (here $L$ is the bolometric luminosity
$L_{\rm bol}$, not the radio luminosity).  The thick solid lines in Fig.~1
show their best fit at different $z$.  We use their best fit to the
bolometric LF at $z=0.1$ to estimate the power per unit volume radiated by
AGN in the local Universe:

\begin{equation}
\rho_{\rm rad} =\int_0^\infty L_{\rm bol}{\phi(L_{\rm bol})\over{\rm ln}10}{{\rm d}L_{\rm bol}\over L_{\rm bol}}.
\label{hop}
\end{equation}

\noindent We find that $\rho_{\rm mech}^{\rm synchro}/\rho_{\rm rad}\simeq
0.018$ (the level of the dashed line in Fig.~2).  The value of $\rho_{\rm
  mech}^{\rm cav}/\rho_{\rm rad}$ depends on the radio luminosity $L_{\rm
  min}$ that is taken for the lower bound of the integration interval in
Eq.~(\ref{blf2}) (solid line in Fig.~2).  For $L_{\rm min}/L_*\simeq 10^{-6}$, 
$\rho_{\rm mech}^{\rm cav}/\rho_{\rm rad}\simeq 0.2$.  For
$\rho_{\rm mech}^{\rm cav}$ to be comparable to $\rho_{\rm rad}$, the
power law of the radio LF would have to extrapolate down to $L_{\rm
  1.4\,GHz}=10^{15}{\rm W\,Hz}^{-1}$ (Fig.~2), which is well below the
allowed limits.

\begin{figure}
\noindent
\centerline{\hbox{\psfig{figure=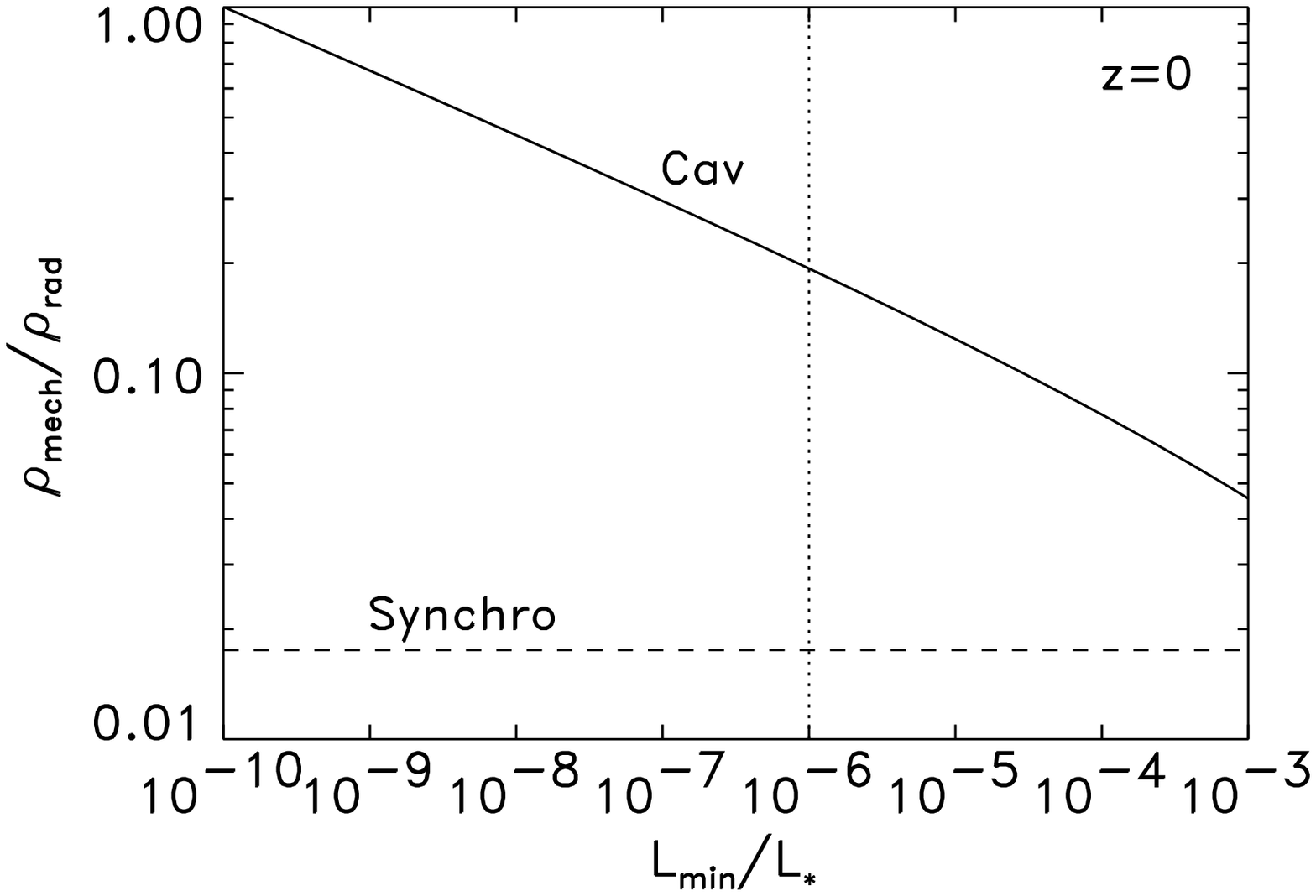,height=6.cm,angle=0}}}
\caption{The ratio between the mechanical power of radio sources
(Eq.~\ref{blf2}) and the radiative power of luminous AGN (Eq.~\ref{hop})
in a representative volume of the local Universe.  The solid line is
computed using the mechanical power estimated from the size of the X-ray
cavities (Eq.~\ref{b2}), in which case $\rho_{\rm mech}$ depends on the
minimum luminosity $L_{\rm min}$ at which one can extrapolate the LF in Eq.~(\ref{blf}).
The dashed line is computed using the mechanical power estimated from the
minimum energy that is needed to explain the observed synchrotron emission
(Eq.~\ref{w2}), in which case $\rho_{\rm mech}$ does not depend on this
uncertainty.  Both lines are computed using the local LF of
\citet{best_etal05}. The vertical dotted line is a conservative lower limit for $L_{\rm min}$.}
\label{fig2}
\end{figure}

\section{The evolution of the mechanical LF}

To go beyond the local Universe, we need to know the evolution of the
radio LF with redshift (assuming that there is no redshift dependence in
Eqs.~\ref{w2} and~\ref{b2}).  We use the radio LFs determined by
\citet{dunlop_peacock90} and \citet{willot_etal01}.

\citet{dunlop_peacock90} modelled the 2.7$\,$GHz LF with the sum of three
contributions: steep-spectrum radio sources in early-type galaxies,
flat-spectrum radio sources in early-type galaxies, and radio sources in
late-type galaxies. The latter are mainly powered by star formation and
are thus irrelevant for our analysis.  Flat-spectrum radio sources are
beamed (the jets are aligned with the line of sight); therefore, the
equations in Section~2 overestimate their mechanical luminosity.  The
addition of the LF of flat-spectrum radio sources to that of
steep-spectrum radio sources makes little difference to the latter, and so
it is safe to deal with this complication by considering only
steep-spectrum radio sources.  The 2.7$\,$GHz LF of steep-spectrum radio
sources in early-type galaxies computed by \citet{dunlop_peacock90} is a
double power-law function of the form of Eq.~(\ref{blf}). In their pure
luminosity evolution model (their other models give similar LFs out to
$z\sim 2$) the dependence of redshift is entirely contained in
$L_*=10^{24.89+1.26z-0.26z^2}{\rm W\,Hz}^{-1}{\rm sr}^{-1}$.  We take this
LF, convert it into a 1.4$\,$GHz LF by assuming an $\alpha=-0.8$ spectral
index, and correct for a cosmology with $\Omega_M=0.3$,
$\Omega_\Lambda=0.7$ and $h=0.7$ so that we can compare our results with
the bolometric LF determined by Hopkins et al. (2007), since
\citet{dunlop_peacock90} had assumed
$\Omega_M=1$, $\Omega_\Lambda=0$ and $h=0.5$.  The blue and red solid
lines in Fig.~1 are obtained by convolving the 1.4$\,$GHz LF determined in
this manner with Eqs.~\ref{w2} and~\ref{b2}, respectively.  Both the blue
and the red curves are significantly below the bolometric LF estimated by
Hopkins et al. (2007; thick solid lines) at all redshifts.

To check this result with an independent determination of the radio LF, we
consider \citet{willot_etal01}'s best fit to the 151$\,$MHz LF (their
model C). We transform it into a 1.4$\,$GHz LF by assuming $\alpha=-0.8$
and correct for the cosmology (Willot et al. 2001 had assumed the same cosmology
as Dunlop \& Peacock 1990).  The blue and the red dashed lines in
Fig.~1 are obtained by convolving the 1.4$\,$GHz LF determined in this
manner with Eqs.~\ref{w2} and~\ref{b2}, respectively. The dashed lines and
the thin solid lines of the same colour run very close to each other.
This demonstrates that the radio LF is not a major source of uncertainty
when it comes to determining the mechanical LF of AGN, even at high
redshifts. 

The mechanical power per unit volume obtained integrating Eq.~(\ref{w2})
over the LF of \citet{dunlop_peacock90}, $\rho_{\rm mech}^{\rm synchro}$,
is shown as a function of redshift in Fig.~3 (dashed line) and compared to
the radiative power per unit volume from Eq.~(\ref{hop}) (symbols). The
dotted line in Fig.~3 is simply $57\rho_{\rm mech}^{\rm synchro}$.  It
shows that the cosmic evolution of the mechanical power density traces
that of the radiative power density, at least for Eq.~(\ref{w2}).  It also
shows that throughout cosmic time jets contribute to a small fraction
($\sim 2\%$) of the AGN energy budget.  
This fraction goes up by a factor of a few at low redshifts,
  if we believe that Eq.~(\ref{b2}) is a more accurate determination of $L_{\rm mech}$
  (solid line in Fig.~3). Even in that case, it is unlikely that the
  mechanical energy accounts for much more than $\sim 10\%$ of the AGN
  cosmic energy budget locally. In addition, the cosmic evolution of the
  jet mechanical energy estimated by Eq.~(4) is weaker than that of the
  radiative power, and so in this case the fraction of the AGN energy
  budget associated with jet mechanical power falls with increasing
  redshift, being well below $10\%$ at $z \gsim 1$.

\begin{figure}
\noindent
\centerline{\hbox{\psfig{figure=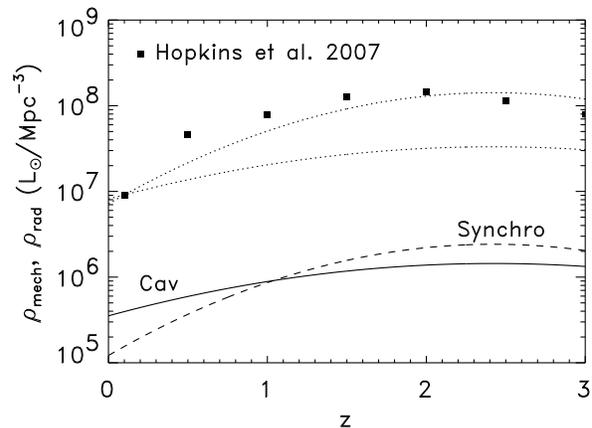,height=6.cm,angle=0}}}
\caption{The mechanical (lines) and radiative (symbols) power of AGN
per unit cosmic volume as a function of redshift.  The mechanical power
per unit volume $\rho_{\rm mech}^{\rm synchro}$ (dashed line) is computed
by inserting Eq.~(\ref{w2}) into Eq.~(\ref{blf2}) and using the radio LF of \citet{dunlop_peacock90}. The solid line shows the redshift dependence of $\rho_{\rm mech}^{\rm cav}$. The vertical normalisation of this line
depends upon the lower luminosity limit of the radio LF adopted
for the integration interval in Eq.~(\ref{blf2}), but the redshift dependence does not. The plotted solid line uses a normalisation that corresponds to $L_{\rm min}=10^{22}{\rm\,W\,Hz}^{-1}$ at $z=0$.
The radiative power per unit volume $\rho_{\rm rad}$ is computed from Eq.~(\ref{hop}) using the bolometric LF of \citet{hopkins_etal07}. The dotted lines correspond to $57\rho_{\rm mech}^{\rm synchro}$ and $19\rho_{\rm mech}^{\rm cav}$.
They are added to ease the comparison of the dashed line and the solid line with the symbols.}
\label{fig3}
\end{figure} 


\section{Discussion and conclusion}
\label{sec:concl} 

Fig.~1 suggests that the radio LF is not a major source of uncertainty
when it comes to determining the mechanical LF of radio sources. It should
be cautioned that constraints on the evolution of the faint end of the
radio LF beyond $z \sim 1$ remain quite poor. The uncertainties are
almost certainly larger than the variations between the different
determinations of the radio LF evolution suggest. Nevertheless, it is also
clear that the main source of uncertainty is the exponent of the $L_{\rm
mech}\propto L_{\rm radio}^\beta$ relation.

The argument based on X-ray cavities suggests that the importance of the
mechanical-energy output decreases in luminous high-accretion-rate
objects.  On the contrary, the argument based on the minimum energy to
produce the observed radio emission suggests that the mechanical-energy
output traces the luminous output. It is not surprising that the latter
conclusion follows from Eq.~(\ref{w2}) since \citet{willot_etal01}, from
which Eqs.~(\ref{w1}) and~(\ref{w2}) are derived, find that the jet
mechanical luminosity is proportional to the optical narrow-line
luminosity. We speculate that the results obtained with the two methods
may be different because the jets in powerful radio sources advance
supersonically and often pierce through the ambient hot gas.  In that
case, the cavities are not inflated gently, therefore the work done to
inflate the cavities is much larger than $pV$. Important progress would
come from observational studies comparing the mechanical luminosities
derived with the two methods for the same objects.

Independently of whether one favours Eq.~(\ref{w2}) or Eq.~(\ref{b2}),
Fig.~1 shows clearly that the bolometric luminosity function is larger
than the mechanical luminosity function by at least one order of magnitude
at all luminosities.  Eq.~(\ref{w2}) establishes a $\sim 2\%$ value
at all redshifts for the mechanical contribution to the AGN cosmic energy
budget (Fig.~3).  Eq.~(\ref{b2}) gives a higher value, which depends on
the minimum luminosity at which the radio LF levels off (Fig.~2).  While
this luminosity is uncertain, we can reasonably estimate that the
contribution inferred from Eq.~(\ref{b2}) is larger than the contribution
inferred from Eq.~(\ref{w2}) by a factor of $\sim 3-10$.  It is thus
unlikely that mechanical energy accounts for much more than $10\%$ of the
AGN cosmic energy budget with $20\%$ as a firm upper limit.  
This is in broad agreement
with previous studies \citep{merloni_08,shankar_etal08a}, which is
encouraging given the different methods that we have used in our
analyses. 

This result implies that radiatively inefficient accretion is unlikely
to contribute to much more than $10\%$ of the BH mass in the Universe,
unless the overall energy efficiency of accretion in this mode is
substantially lower than the energy efficiency 
in the radiatively efficient mode,
ie unless a substantial fraction of the energy is advected onto the
BH, rather than coming out as either photons or jets.
Based on an ADAF (advection dominated accretion flow; \citealp{narayan_yi94}) 
model,
\citet{merloni_08} suggested a kinetic efficiency for the production of jets
of $\epsilon_k \simeq 0.005$ (where $L_{\rm mech} = \epsilon_k
\dot{M}_\bullet{\rm c}^2$) in low-accretion rate AGN,
and thus concluded that $\sim 18-27\%$ of the
BH growth occurs in a radio-jet-producing mode. 
\citet{shankar_etal08a} found a similar value ($\sim 20-30\%$) by
deriving a kinetic efficiency of $\epsilon_k\sim 0.01$ 
for the production of radio jets in radiatively efficient
radio-loud AGN, and by adopting this value for all radio sources. 
However, it is not clear whether such a low value for $\epsilon_k$ is
appropriate for radiatively inefficient AGN, which may channel
most of the accretion power into jets \citep{blandford_begelman99}.

How do our results fit in the emergent scenario, in which jet heating
plays a major role in the evolution of early-type galaxies and galaxy
clusters?  The energy released by the formation of a supermassive BHs is
two orders of magnitude larger than the host galaxy's binding energy, so
the issue is not the energy, but the efficiency with which it can absorbed
by the ambient gas.  Photoionisation of the inner orbitals of metals and
Compton scattering are the main processes by which AGN radiation heats the
surrounding gas.  In nearby massive elliptical galaxies such as the
systems studied by \citet{allen_etal06}, the gas on a galactic and group
or cluster scale is hot and highly transparent, having $n_H\sim
10^{21}{\rm\,cm}^{-2}$.  This column density is only a thousandth the
column density $\sigma_{\rm T}^{-1}$ above which the gas becomes Thomson
thick ($\sigma_{\rm T}$ is Thomson cross section for electron scattering),
meaning that only $\sim 1$ photon in $10^3$ is scattered by an electron
before leaving the galaxy. Even discounting any inefficiencies in the
transfer of the photon energy to the gas in the scattering process
(ie Compton scattering transfers ${h\nu\over m_{\rm e}{\rm c}^2}<1\%$ of the photon energy to a free electron 
per scattering event), 
the fraction of the luminous energy that would be absorbed by the gas if a
quasar switched on in a nearby massive elliptical galaxy would be $\lsim
0.1\%$.  Therefore, based on this argument, mechanical heating could be $>
10$ times more important than radiative heating even if the mechanical
power were $100$ times smaller than the radiative power.

There is broad observational evidence that mechanical heating by jets
plays an important role in solving the cooling-flow problem in galaxy
clusters.  The evidence that the same solution can be applied to
individual galaxies is much weaker because jets may be collimated on
kiloparsec scales and transport most of the energy to beyond the gaseous
halo of the host galaxy. However, the bulk of the jet mechanical power is
produced in low luminosity radio sources, most of which have small radio
sizes. Even in larger sources, jet-interstellar medium interactions may
also occur on sub-kiloparsec scales (the knots in the jets of M87 may be
evidence for this). Therefore, in individual ellipticals
 the role of weak radio sources vs. episodic
quasar heating (eg  \citealp{ciotti_ostriker07}) remains an open problem.


\section*{Acknowledgments}

A.C. acknowledges discussion with A. Merloni.
  

\bibliographystyle{mn2e}

\bibliography{ref_av}

\label{lastpage}
\end{document}